\documentstyle[aps,psfig,twocolumn]{revtex}
\begin{document}
\draft 
\input amssym.def
\input amssym.tex
%
\def\R{{\Bbb R}}
\def\Z{{\Bbb Z}}
\def\ds{\displaystyle}
\def\e{\varepsilon}
\def\cmls{coupled map lattices}
\def\Cml{Coupled map lattice}
\def\cml{coupled map lattice}
\def\PLA{Phys.~Lett.~A}
\def\PRL{Phys.~Rev.~Lett.}
\def\PRA{Phys.~Rev.~A}
\def\PRE{Phys.~Rev.~E}
\def\PD{Physica D}
\def\IJBC{Int.~J.~Bifurcation and Chaos}
\def\PTRSL{Phil.~Trans.~R.~Soc.~Lond.}
\def\PRB{Phys.~Rev.~B}
\def\JTB{J.~theo.~Biol.}
\def\JAE{J.~Anim.~Ecology}
\def\PTP{Prog.~Theor.~Phys.}
\def\Kan{K.~Kaneko}
\def\RCG{R.~Carretero-Gonz\'alez {\em et.al}.}
\def\FigSize{8.40}\def\FigJump{0.25cm}     
%
\def\oneFIG#1#2#3{\begin{figure}[ht]
\centerline{\psfig{figure=#1,width=#2cm,silent=}}
\vskip \FigJump \caption[]{#3\label{#1}} \end{figure} }
%
\def\twoFIG#1#2#3#4{\begin{figure}[ht]
\centerline{\psfig{figure=#1,width=#3cm,silent=}}
\vskip 0.2cm
\centerline{\psfig{figure=#2,width=#3cm,silent=}}
\vskip \FigJump \caption[]{#4\label{#1}} \end{figure} }
%
\def\PULLINGCAP
{The dynamics of a front for a one-way CML results from the
competition between local dynamics and coupling.
The schematic contributions from the local dynamics (arrows 
with filled arrow-head) and coupling (arrows with empty arrow-head)
are depicted for all front sites at time $t$ (filled circles).
A sufficiently large coupling causes a site located
within the basin $I_+$ (filled circle at the centre of front)
to switch to the basin $I_-$, and move 
rapidly towards $x_-^*$. As a consequence, the centre of mass 
of the front will move to the right, resulting in propagation.
\par}
\def\PHASPACAP
{Phase space $h'(t)$ vs.~$h(t)$ of the ODEs (\ref{ODEs}) corresponding 
to the travelling front solution in the continuum limit. 
(a) A one-way CML corresponds to Hamiltonian motions. 
(b) A diffusive CML corresponds to dissipative motions. 
Note that in (b) an heteroclinic connection between
unstable points can still exist in presence of friction.\par}
\def\WAVESCAP
{Travelling front solutions in the continuum limit approach. 
(a) Heteroclinic ($(i)$ and $(ii)$), and oscillatory $(iii)$ solutions 
in a one-way CML.
(b) Damped heteroclinic solutions: $(i)$ connects the stable fixed point 
$x_+^*$ to the unstable fixed point $x^*=p$; $(ii)$ connects
the two stable points.\par}
\def\SFTCAP
{The travelling front shape is reconstructed by superimposing 
snapshots of the discrete interface in a co-moving reference frame. 
(a) One-way coupling: $f(x)=\tanh(x/0.2)$, $\e=0.398011$, 
$v(\e)\simeq 0.2856031\simeq 2/7$.
(b) Diffusive coupling: $f$ is the second iterate of the logistic 
map, with $\e=0.2$ and $v(\e)\simeq 0.0097915$; 
(c) Diffusive coupling: $f$ as in (b), with $\e=0.6$ and
$v(\e)\simeq 0.1118273$.\par}
\def\BOUNDARYCAP
{Approximating travelling front for a rational velocity. 
The parametric point located at the edge of a tongue (small box), 
is approached both transversally (path $A$) and tangentially 
(paths $B$ and $C$). 
The tongue corresponds to a travelling front with velocity 
$v(\e)=1/3$, which is periodic with period 3. 
In all cases the front shape approaches the same step 
function, with 3 steps per unit length in $z$. 
Note that the fronts have been shifted for clarity.
Here the local map is $f(x)=\tanh(x/\nu)$, and the parameters
at the boundary of the mode-locking region are $(\e_*=0.3983, \nu_*=0.1)$. 
\par}
\def\AUXMAPCAP
{Auxiliary maps $\Phi$ for the central site of the interface
defined in the square region depicted by the thick lines. 
(a) one-way CML: local map $f(x)=\tanh(x/0.2)$ with
$\e=0.4$, $v=0.28973453$.
(b) Diffusive: same parameters as in Figure \ref{sftG.ps}(c). 
The delay Poincar\'e section $\Phi(x)$ corresponds to the central 
rectangular region of each plot (region 1) in Figure (a). 
Each rectangular region corresponds to the return map for 
a particular combination of sites. For instance, the region 2)
in Figure (a)  corresponds to $\bar x_{t+1}(1)$  vs.~$\bar x_t(1)$.
\par}
\def\AUXMAPTONGUECAP
{Typical reconstruction of the auxiliary map $\Phi$ inside a 
mode-locking tongue.
The large stars locate the original periodic orbit well 
inside a tongue (in this example $v=1/5$). A small random 
perturbation is periodically applied to the central site of the front.
The state of each perturbed front is depicted by circles. 
After few transient iterations (2 or 3), the perturbed front
relaxes onto the one-dimensional manifold represented by the thick line.
This technique is applied repeatedly until the whole one-dimensional
manifold is filled in. 
\par}
\def\CIRCLECAP
{The auxiliary map $\Phi$, accounting for the dynamics of the 
site closest to the unstable point $x^*$, is a circle map on 
$[x^*-a,x^*+a]$, with two increasing branches $f_+$ and $f_-$.
\par}
\def\INTERMITTENCYCAP
{Onset of intermittent regime in the auxiliary map,
corresponding to the development of a step-like 
travelling front. For the parameter values and the
front shape please refer to Figure \ref{sftG.ps}(c). 
The intermittency is the precursor 
of a pair of period-7 orbits.
\par}
\def\TONGUESCAP
{Principal Arnold's tongues of the travelling front velocity in the
one-way CML with the hyperbolic tangent local map $f(x)=\tanh(x/\nu)$
in the $(\e,\nu)\in[0,1]^2$ parameter space. The right hand side scale
gives the width $\sigma^2$ of the corresponding travelling front
for fixed $\e=1/2$.
\par}
\def\AUXMAPCONTINUUMCAP
{Auxiliary maps $\Phi_i(x)$ for the reduced dynamics of the travelling
front near the continuum limit. The CML was taken to be one-way with 
local map $f(x)=\tanh(x/\nu)$, $\nu=100/101$ and coupling strength 
$\e=0.45$.
\par}

\title{\bf One-dimensional dynamics for travelling fronts in 
           coupled map lattices}

\author{R.~Carretero-Gonz\'alez\thanks{Current address:
	Centre for Nonlinear Dynamics and its Applications,
	University College London,
	Gower Street, LONDON WC1E 6BT, U.K.
	E-mail: R.Carretero@ucl.ac.uk,
        URL: http://www.ucl.ac.uk/$\sim$ucesrca/},
	D.K.~Arrowsmith and F.~Vivaldi\\
   {\sl School of Mathematical Sciences, Queen Mary and Westfield College,}\\
   {\sl Mile End Road, London E1 4NS, U.K.}}
\date{Submitted to Physical Review E, April 1999.}
\maketitle

\hrule\vskip -0.1cm
\begin{abstract}
Multistable coupled map lattices typically support travelling fronts, 
separating two adjacent stable phases.  We show how the existence of an 
invariant function describing the front profile, allows a reduction of 
the infinitely-dimensional dynamics to a one-dimensional circle homeomorphism, 
whose rotation number gives the propagation velocity. The mode-locking 
of the velocity with respect to the system parameters then typically follows. 
We study the behaviour of fronts near the boundary of parametric stability, 
and we explain how the mode-locking tends to disappear as we approach the 
continuum limit of an infinite density of sites.
\end{abstract}
\medskip \hrule


\section[Introduction]{Introduction}
Coupled map lattices (CML) are arrays of low-dim\-en\-sio\-nal dynamical
systems with discrete time, originally introduced in 1984 as
simple models for spatio-temporal complexity \cite{CML:84}.
CMLs have been extensively used in modelling spatio-temporal chaos
in fluids phenomena such as turbulence \cite{Beck:94-Kan:89},
convection \cite{Yanagita:93} and open flows \cite{Willeboordse:95}.
Equally important is the analysis of pattern dynamics, which
has found applications in chemistry \cite{Kapral:94} and
patch population dynamics \cite{Hassell:95-Sole:95}.
One important feature of pattern dynamics is the existence of
travelling fronts, which occur at the pattern boundaries,
and are also seen to emerge from apparently decorrelated
media \cite{Kan:92-93}.
This paper extends the work on the behaviour of a travelling
interface on a lattice developed in
\cite{rcg:thesis,rcg:modloc,rcg:lowdim,Coutinho:98}.
Our main results are: 
$i)$ a constructive procedure for the reduction of the 
infinitely-dimensional dynamics of a front to one dimension;
$ii)$ a characterization of the behaviour of fronts near the 
boundary of parametric stability;
$iii)$ a characterization of the behaviour of fronts near the 
continuum limit.

We consider a one-dimensional infinite array of sites.
At the $i$-th site there is a real dynamical variable $x(i)$,
and a local dynamical system ---the {\em local map}.
The latter is given by a real function $f$ which we assume to be
the same at all sites.
The dynamics of the CML is a combination of local dynamics and
coupling, which consists of a weighted sum over some
neighbourhood.
The time-evolution of the $i$-th variable is given by
$$
x_{t+1}(i) = \sum_k \e_k f(x_t(i+k))
$$
where the range of summation defines the neighbourhood.
The coupling parameters $\e_k$ are site-independent,
and they satisfy the conservation law $\sum \e_k=1$,
to prevent unboundedness as time increases to infinity.
The two most common choices for the coupling are
\begin{equation} \label{one-way}
x_{t+1}(i)= (1\!-\!\e) f(x_t(i)) + \e f(x_t(i\!-\!1)),
\end{equation}
and
\begin{equation} \label{diffu}
x_{t+1}(i)  =  (1\!-\!\e) f(x_t(i))
	  \ds + {\e\over 2}\left( f(x_t(i\!-\!1)) + f(x_t(i\!+\!1)) \right)
\end{equation}
which are called {\em one-way\/} and {\em diffusive\/} CML, respectively.
The diffusive CML corresponds to the discrete analogue of the
reaction-diffusion equation with a symmetrical neighbouring interaction.
There is now a single coupling parameter $\e$ which is constrained
by the inequality $0\leq\e\leq1$, to ensure that the sign of the
coupling coefficients in (\ref{diffu}) and (\ref{one-way})
({\em i.e.}~$\e$, $\e/2$ and $1-\e$) remains positive.

In this paper we study front propagation in {\em bistable\/} CMLs.
The local mapping $f$ is continuous and has two stable equilibria, 
and a {\em front\/} is any monotonic arrangement of the
state variables, linking asymptotically the two equilibria.

We will show how to construct a one-dimensional circle map
describing the motion of the front. Such a mapping
originates from the existence of an invariant function 
describing the asymptotic front profile, and of a one-dimensional 
manifold supporting the transient motions.
The rotation number of the circle map will then give the velocity 
of propagation, resulting in the occurrence of {\em mode-locking}, 
{\em i.e.}, the parametric stability of the configurations 
that correspond to {\em rational\/} velocity. 
We will describe the vanishing of this phenomenon in the 
continuum limit, as the width of the front becomes infinite.
We shall also be concerned with the evolution of the front
shape near the boundary of parametric stability, where the 
continuity of the local map ensures a smooth evolution
of the front shape.

\newpage

Velocity mode-locking is commonplace in nonlinear coupled systems
({\em e.g.}, Frenkel-Kontorova models \cite{Floria:96}, Josephson-junction
arrays \cite{Basler:97}, excitable chemically reactions
\cite{Schreiber:94}, and nonlinear oscillators 
\cite{Bressloff:97-Kuske:97}): the present work provides 
further support for its genericity, and highlights key
dynamical aspects.

Throughout this paper, the very existence of fronts in the
regimes of interest to us is inferred from extensive
numerical evidence. We are not concerned with existence 
proofs here. Fronts have been proved to exist in various
situations, mainly for {\em discontinuous\/} piecewise 
affine maps (see \cite{Coutinho:98} and references therein);
in the present context however, continuity is crucial.

Following \cite{rcg:modloc}, we consider a CML whose local map
$f$ is continuous, monotonically increasing and which possesses
exactly two stable fixed points ${x^*_-}$ and ${x^*_+}$.
It then follows that there exists a unique unstable fixed point
$x^*$ such that $x_-^*<x^*<x_+^*$.
The homogeneous fixed states $x(i)=x_\pm^*$,
$\forall\, i \in \Bbb Z$, inherit the stability of the fixed points
$x_\pm^*$ \cite{Gade:93-Zhilin:94b}.
We denote by $I_-=[x_-^*,x^*)$ and $I_+=(x^*,x_+^*]$
the basins of attraction of $x_-^*$ and $x_+^*$, respectively,
while $I=[x^*_-,x^*_+]$.

A {\em minimal mass state\/} is a state satisfying the
monotonicity condition $x(i)\leq x(i+1)$, for all $i$.
It can be shown directly from the system equation that the image
of a minimal mass state has the same property.
A {\em front\/} is a minimal mass state satisfying the
asymptotic condition:\/ $\lim_{i\to\pm\infty}x(i)=x^*_\pm.$
The main properties of a front are its {\em centre of mass\/} $\mu_t$
and its {\em width\/} $\sigma_t^2$, which measure its position and
spread at time $t$, respectively.
They are defined as the mean and variance of the variable $i$
with respect to the time-dependent probability distribution
\begin{equation}\label{prob}
\ds p_t(i)={|\Delta x_t(i)|\over\ds\sum_{i=-\infty}^{\infty}{|\Delta x_t(i)}|},
\end{equation}
where $\Delta x_t(i)=x_t(i+1)-x_t(i)$ is the variation of the local states.
We have
\begin{equation} \label{mu_sigma}
\ds\begin{array}{rcl}
\mu_t & = & \ds\sum_{i=-\infty}^{\infty}{i p_t(i)},\\[3.0ex]
\sigma_t^2 & = & \ds\sum_{i=-\infty}^{\infty}{(i-\mu_t)^2 p_t(i)}.
\end{array}
\end{equation}
A state $X_t=\{x_t(i)\}$ with finite centre of mass and width is said
to be {\em localised}.

In this paper we are interested in fronts of {\em fixed shape}, 
moving at velocity $v$. They are described by the equation
\begin{equation}\label{eq:Front}
x_t(i)=h(i-vt);
\qquad
v=\lim_{t\to\infty}{\mu_t\over t},
\qquad  t, i\in{\Bbb Z}.
\end{equation}
Here the function $h:\R \mapsto [x^*_-,x^*_+]=I$ is to 
be determined subject 
to the condition that it be monotonic, with 
$\lim_{x\to\pm\infty}=\pm x^*_\pm$. 
The degree of smoothness of $h$ will depend on the regime being considered.

The object of interest to us is the central part of the front.
Far away from the centre, the lattice is almost homogeneous
({\em i.e.}, $|\Delta x_t(i)|\ll |I|$), and the dynamics is dominated by 
the attraction towards the stable points of the local map.
The qualitative evolution of the centre of the front can be 
understood as the result of the competition between local 
dynamics and coupling (see Figure \ref{pulling.eps}, for the one-way case).
For small $\e$, the attraction towards the fixed points $x_\pm^*$
overcomes the effect of the coupling, resulting in propagation
failure (zero velocity) \cite{rcg:modloc}.
A sufficiently large coupling will instead cause a site located
within the basin $I_+$ to switch to the basin $I_-$, and move 
rapidly towards $x_-^*$. As a consequence, the centre of mass 
of the front will move to the right, resulting in propagation.

\oneFIG{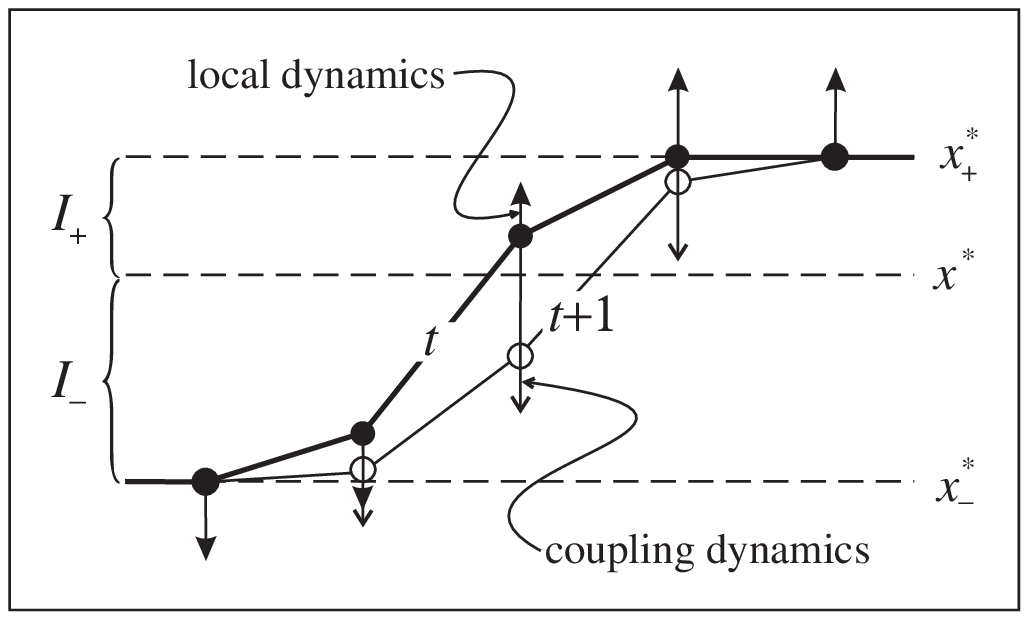}{\FigSize}{\PULLINGCAP}

A similar argument can be applied in the diffusive case.
Now however the coupling is symmetric, and a bias to either 
of the stable points will have to be introduced via an
asymmetry in the local map.
For instance, increasing the size of the basin of attraction of
$x_-^*$, will result in propagation from left to right for
an increasing front.

In previous works we have shown that the dynamics of a
{\em finite-size\/} interface in a class of piece-wise linear
one-way CMLs can be reduced to a single one-dimensional map
\cite{rcg:modloc,rcg:lowdim}.
The finiteness of the front depended on the existence of
degenerate superstable fixed points of the local map,
that caused nearby orbits to collapse unto the stable states
in a single iteration.
In this paper we remove such degeneracy, and consider smooth
local maps and infinitely extended fronts (the case of a
discontinuous local map was treated in \cite{Coutinho:98}).
We shall provide evidence that every front evolves towards a 
unique asymptotic regime, characterized by a constant velocity 
as well as an invariant shape.
Under these assumptions, we then show how the front behaves 
at the boundary of the regions of parametric stability 
(here the continuity of the local map is essential), 
and how the reduction to one-dimensional dynamics can 
be achieved.

This paper is organised as follows.
In section \ref{SEC:CONTINUUM} we describe the behaviour of 
travelling fronts in the continuum limit, when the density of 
interfacial sites is large.
We obtain an ODE describing the shape of the travelling front,
and with it we find new classes of fronts.
%
In section \ref{SEC:REDUCED} we consider the asymptotic 
shape of the front, and we provide extensive evidence that
such a shape is fixed and is described by a continuous function.
This result allows us to derive a procedure for the reduction of
the infinite-dimensional interface dynamics to a one-dimensional
problem described by the {\em auxiliary map.}
%
In section \ref{SEC:MODLOC} we show that the auxiliary map
is a circle map and we relate its rotation number to the velocity 
of the front, from which the mode-locking of the velocity with 
respect to the system parameters follows. 
Finally, we explain in terms of reduced dynamics the
vanishing effect of mode-locking when the continuum
limit is approached.
\bigskip

\section[The continuum limit]
{The continuum limit{\label{SEC:CONTINUUM}}}
In this section we consider fronts with large widths,
for which the relative density of sites is large, and 
the continuum approximation becomes appropriate.
To achieve a front with such features, the attraction 
towards $x_\pm^*$ and the repulsion of $x^*$ must be small. 
Because $f$ is continuous and monotonic, then $f$ is 
necessarily close to the identity, {\em i.e.} 
$$
\delta_f=\sup_{x^*_-<x<x^*_+}\,|f(x)-x| \,\ll\, 1.
$$
Choosing functions $f$ such that $\delta_f\rightarrow 0$ 
is referred to as the {\em continuum limit}.

Inserting equation (\ref{eq:Front}) into the equations of motion
(\ref{one-way}) and (\ref{diffu})
we find that
\begin{equation}\label{functional}\begin{array}{lrcl}
\hbox{\rm a)}& h(z-v)&=&(1-\e)f(h(z))+ \e f(h(z-1)),\\[3.0ex]
\hbox{\rm b)}& h(z-v)&=&(1-\e)f(h(z))\\[1.0ex]
             &       & &+ \ds{\e\over 2}\left(f(h(z-1))+f(h(z+1))\right),
\end{array}\end{equation}
for the one-way and diffusive CML, respectively, 
where $z=i-vt$.
A function $h$ satisfying the functional equation (\ref{functional})
represents the fixed shape of a front travelling at the velocity $v$.

To solve equation (\ref{functional}) in the continuum limit,
we assume $f$ and $h$ to be twice differentiable, and consider 
the Taylor series of $h$ in $z$, up to second order.
The Taylor expansion becomes accurate as the width increases, 
since in this case the variation of $h$ over adjacent lattice 
sites tends to zero.
We obtain
\begin{equation}\label{taylor}
\begin{array}{rl}
h(z)-f(h(z))& + A\, h'(z)
-\ds\left({{\ds\e\,f''(h(z))}\over\ds 2}\right) {h'(z)}^2 \\[2.0ex]
&+\ds\left( {\ds v^2-\e\,f'(h(z))\over\ds 2}\right) h''(z)\,=\,0,
\end{array}
\end{equation}
where $A=\left(\e\,f'(h(z))- v\right)$  and $A=-v$, for the one-way
and diffusive CML, respectively. In the continuum limit we can
further simplify equation (\ref{taylor}) by considering $f'(x)=1$
and $f''(x)=0$, to obtain
\begin{equation}\label{ODEs}\begin{array}{ll}
\hbox{\rm a)}&
 h(z)-f(h(z))+\ds \left({\ds\e(\e-1)\over\ds 2}\right) h''(z)=0, \\[4.0ex]
\hbox{\rm b)}&
 h(z)-f(h(z)) - v\,h'(z) + \ds
    \left({\ds v^2-\e\over\ds 2}\right) h''(z)=0,
\end{array}\end{equation}
for the one-way and diffusive CML, respectively, 
where we set $v=\e$ in the one-way case since in the continuum limit
$f(x)\rightarrow x$ and thus the rate of information exchange
({\em i.e.}~the velocity) is equal to $\e$.
For the diffusive case the velocity is not equal to $\e$, since the
total information exchange comes from the competition between the
left and right neighbours.
Nevertheless, as we shall see, it is possible to give an analytical
approximation to the velocity for the case of an asymmetric cubic 
local map.

\oneFIG{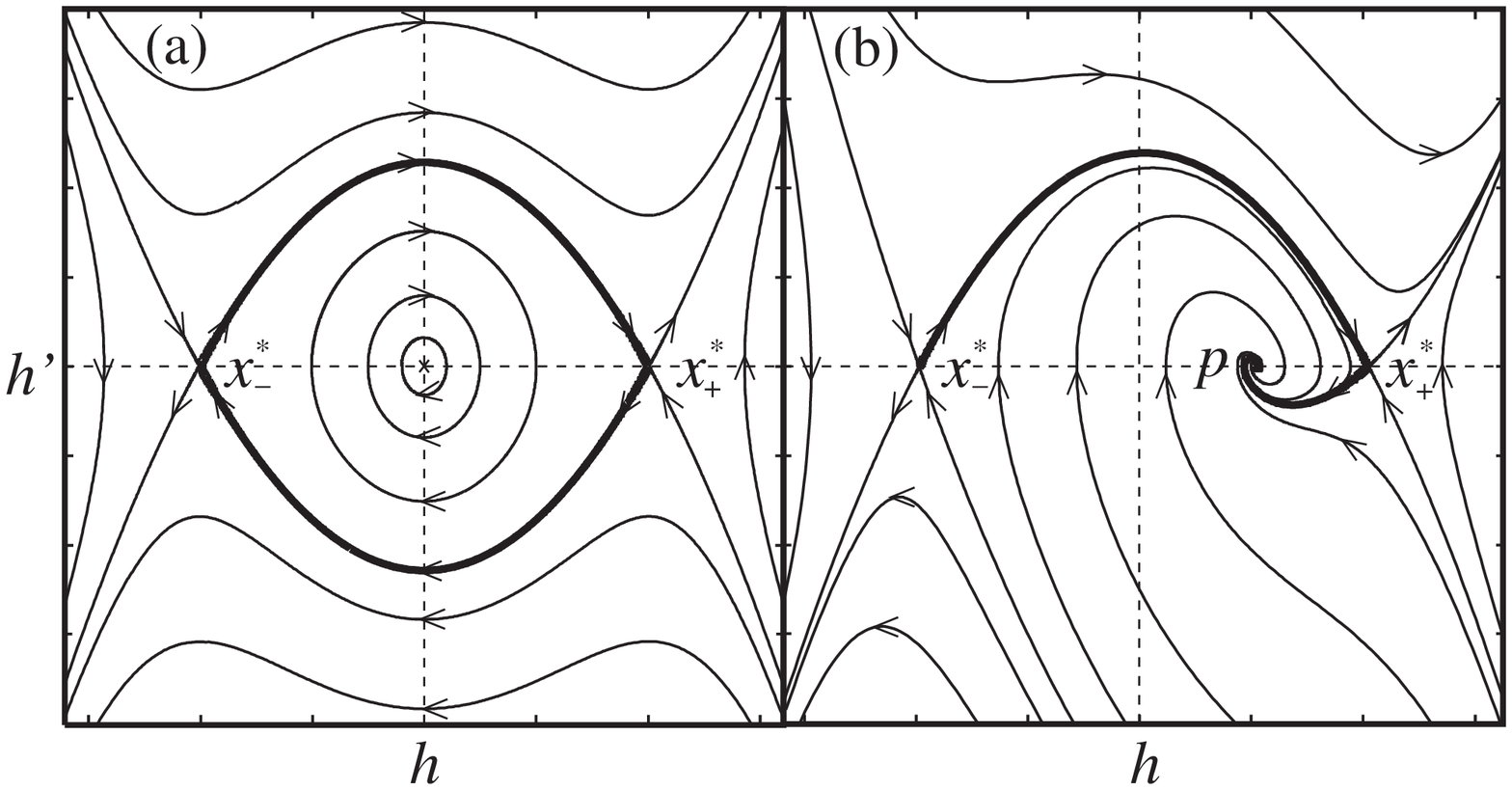}{\FigSize}{\PHASPACAP}

Equations (\ref{ODEs}) are similar to those obtained in
\cite{Chow:95-96}, where the travelling front in
a lattice of coupled ODEs, is reduced to a single equation.
The ODEs (\ref{ODEs}) describe the motion of a particle of mass
$m=(v^2-\e)/2$, subject to the potential $V(x)=\int{(f(x)-x)\,dx}$,
with maxima located at the stable fixed points of the local map
(Figure \ref{pha-spa.eps}).

In the one-way case, the system is conservative.
For numerical experiments, we choose a symmetric local map
$f$ with fixed point $x_\pm^*=\pm1$ and $x^*=0$.
The resulting potential is also symmetric.
There exist two heteroclinic connections, joining $x_-^*$
to $x_+^*$, and $x_+^*$ to $x_-^*$, respectively
(the thick lines in Figure \ref{pha-spa.eps}(a)).
They correspond, respectively, to an increasing and a
decreasing symmetric travelling front for the CML.

In the diffusive case, the differential equation has
the dissipative term $-v\,h'(z)$.
For the local map, we choose $0<x^*=p<1$, which introduces
an asymmetry in the system, and the maxima of the potential
are now unequal:\/ $V(x_-^*)>V(x_+^*)$.
Imposing a heteroclinic connection from $x_-^*$ and $x_+^*$,
constrains the velocity $v$ of the front (see below).
For larger velocities, the separatrix emanating from $x_-^*$
approaches $p$, while for smaller $v$ it escapes to infinity.
Since the presence of friction breaks the time-reversal
symmetry, only one heteroclinic connection is possible, and
the separatrix emanating from $x_+^*$ always approaches
$p$ (the thick lines in Figure \ref{pha-spa.eps}(b)).

\oneFIG{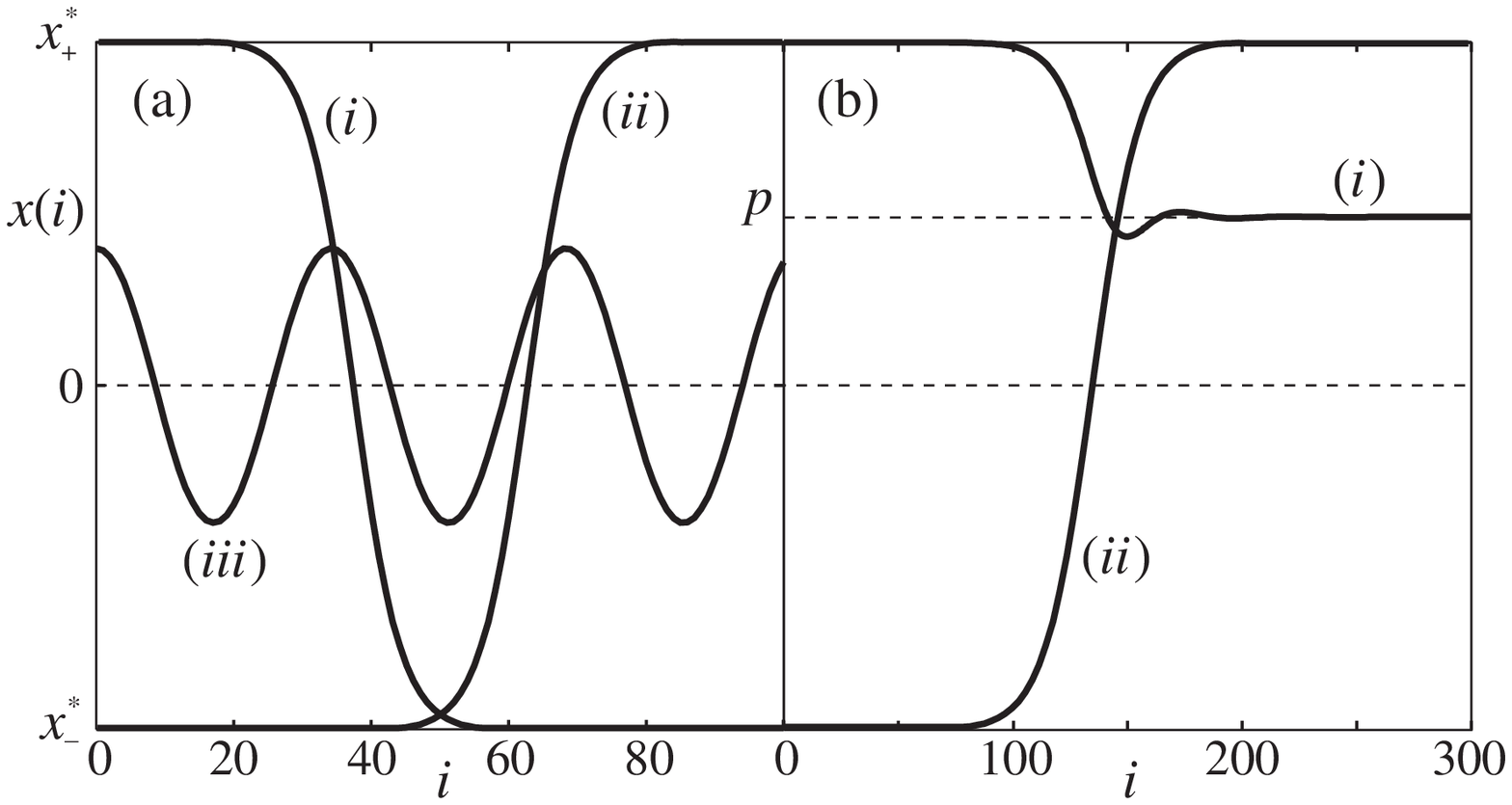}{\FigSize}{\WAVESCAP}

The continuum approximation can be used to construct new 
kinds of travelling fronts. For example, the librating
orbits in Figure \ref{pha-spa.eps}(a) (one-way case), correspond 
to spatially-periodic travelling fronts that never touch the 
stable points (see Figure \ref{waves.eps}(a) $(iii)$).
Such spatially-periodic orbits do not exist in the diffusive case. 
Nevertheless, it is possible to construct the travelling front
departing from $x_+^*$ that dissipates down to $p$.
This new solution has a damped oscillatory profile (see Figure
\ref{waves.eps}(b) $(i)$), and it is unstable, because 
the fixed point at $p$ is unstable (for the CML).

In the remainder of this section, we briefly examine the case of
a cubic local map, providing the dominant behaviour of a general
bistable local map in the continuum limit.
We use the one-parameter families of cubics
\begin{equation}\label{cubics}
\begin{array}{ll}
\hbox{\rm a)} &
f(x)={\displaystyle x\over \strut \displaystyle 2}(3-\nu-(1-\nu)\, x^2)\, ,
	   \\[4.0ex]
\hbox{\rm b)} &
f(x)=(1-\nu)\,(p\,x^2-x^3 -p)+(2-\nu)\,x.
\end{array}\end{equation}
for the one-way and diffusive CML, respectively.
Again, $x_\pm^*=\pm 1$ for both cases, while $x^*=0$ in the
one-way case and $x^*=p$ in the diffusive case, where $0<p<1$
controls the asymmetry.
The continuum limit is attained by letting the parameter
$\nu$ approach 1 from below.
Substituting the cubic local maps (\ref{cubics}) in the differential
equations (\ref{ODEs}), one finds expressions for the heteroclinic
connections corresponding to the travelling front solutions:
\begin{equation}\label{wave-sol}
\begin{array}{ll}
\hbox{\rm a)} &
h(z) = \tanh\left(\sqrt{\ds 1-\nu \over\ds 2\e(1-\e)}\,z\right), \\[4.0ex]
\hbox{\rm b)} &
h(z) = \tanh\left({\ds(1-\nu)p \over\ds v}\,z\right),
\end{array}\end{equation}
where
\begin{equation}\label{width}
\begin{array}{lll}
\hbox{\rm a)} &
v=\e &\quad \ds \sigma^2 = {2\pi^2\over 3}{\e(1-\e)\over 1-\nu} \\[4.0ex]
\hbox{\rm b)} &
v= p\sqrt{\e(1-\nu)} & \quad\ds \sigma^2 = {\pi^2\over 3}{\e \over 1-\nu}.
\end{array}
\end{equation}
for the one-way and diffusive CML, respectively.
In the diffusive case, the expression for the velocity is 
derived from imposing an heteroclinic connection, while 
the scaling of the width $\sigma^2$ is found from the 
solutions (\ref{wave-sol}).
Note that for both models the functional dependence of the width on
the parameter $\nu$ is the same, and it describes the rate at which
the front broadens as the continuum limit is approached.
Moreover, from (\ref{wave-sol}) and (\ref{width}) we have
that in the diffusive case $h$ is independent of $p$.

While in the continuum limit the front is described by a
continuous function $h$ ({\em cf.}~equations (\ref{wave-sol})),
there is no {\em a-priori\/} reason why such a function should continue
to exist away from the limit, due to the discrete nature of the system.
We shall nonetheless give evidence that the dynamics of a
front far from the continuous limit remains one-dimensional.

\section[Reduced dynamics of the travelling front]
{Reduced dynamics of the travelling front{\label{SEC:REDUCED}}}

In this section we provide evidence that every front has a fixed 
profile, which can be characterized by an invariant function $h$.
Such a function will then be used to construct a
one-dimensional mapping describing the front evolution
---the {\em auxiliary map}.

If the velocity $v$ of the front is {\em irrational}, then the 
collection of points $i-vt$, with $i$ and $t$ integers, form a 
set dense on the real line. 
Numerical experiments consistently suggest that 
in the case of a front, the closure of the set of points 
$(i-vt,\,x_t(i))\in\R^2$ forms the graph of a continuous and 
monotonic function: \/ $h:\Z \mapsto [x_-^*,\,x_+^*]$, which
is a solution to the functional equation (\ref{functional}).

The results for both CML models are summarised in Figure \ref{sftG.ps},
where we have superposed all translates of the discrete fronts, 
after eliminating transient behaviour.
This procedure requires computing $v$ numerically, which was 
done using some $10^7$--$10^8$ iterations of the CML.
(In principle, a numerical solution to (\ref{functional}) can 
be found using various iterative functional schemes.
However, all the schemes considered were plagued by slow 
convergence and are not discussed here.)

\oneFIG{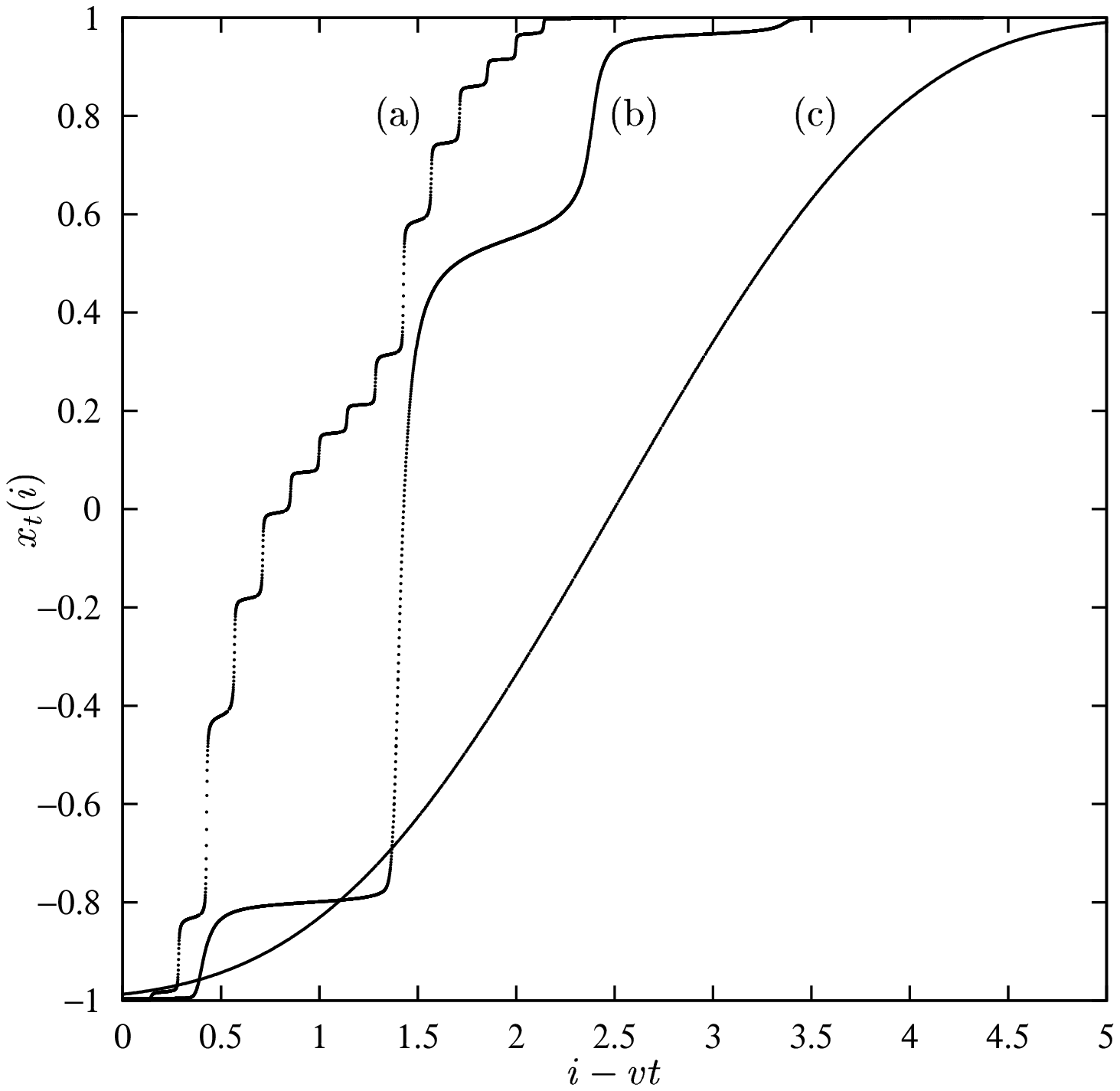}{\FigSize}{\SFTCAP}

In the case in which $v=p/q$ is {\em rational}, the function $h$ 
is specified only at a set of $q$ equally spaced points.
It turns out, however, that the definition of $h$ becomes 
unequivocal in a prominent parametric regime, corresponding 
to the boundary of the so-called mode-locking region or 
{\em tongue}. The latter is defined as the collection of 
parameters $(\e,\nu)$ corresponding to a given rational velocity,
where $\nu$ (not necessarily one-dimensional) parametrizes the 
family of local maps ---for the one-way CML we typically use 
$f(x)=\tanh(x/\nu)$.

We defer the discussion of the origin of such regions to the next section. 
Here we consider a sequence of parameters 
$(\e_n,\nu_n)\rightarrow (\e_*,\nu_*)$, 
converging from the outside towards a boundary point 
$(\e_*,\nu_*)$ of the tongue (see Figure \ref{boundary.ps}). 
Independently from the path chosen to approach the boundary 
point, the front $h$ appears to approach a unique limiting 
shape. The limiting shape is a step function with $q$ steps (where $v=p/q$)
for every unit length ---the horizontal length of each step is $1/q$
since there are $q$ equidistant points in every horizontal interval
of unit length for a $v=p/q$ orbit.
In the limit, the front dynamics becomes periodic, with 
periodic points corresponding to the midpoint of each step.
This observation suggests that choosing step fronts with the 
periodic points at their midpoints ensures continuity of the 
front shapes at the resonance tongue boundaries.

\oneFIG{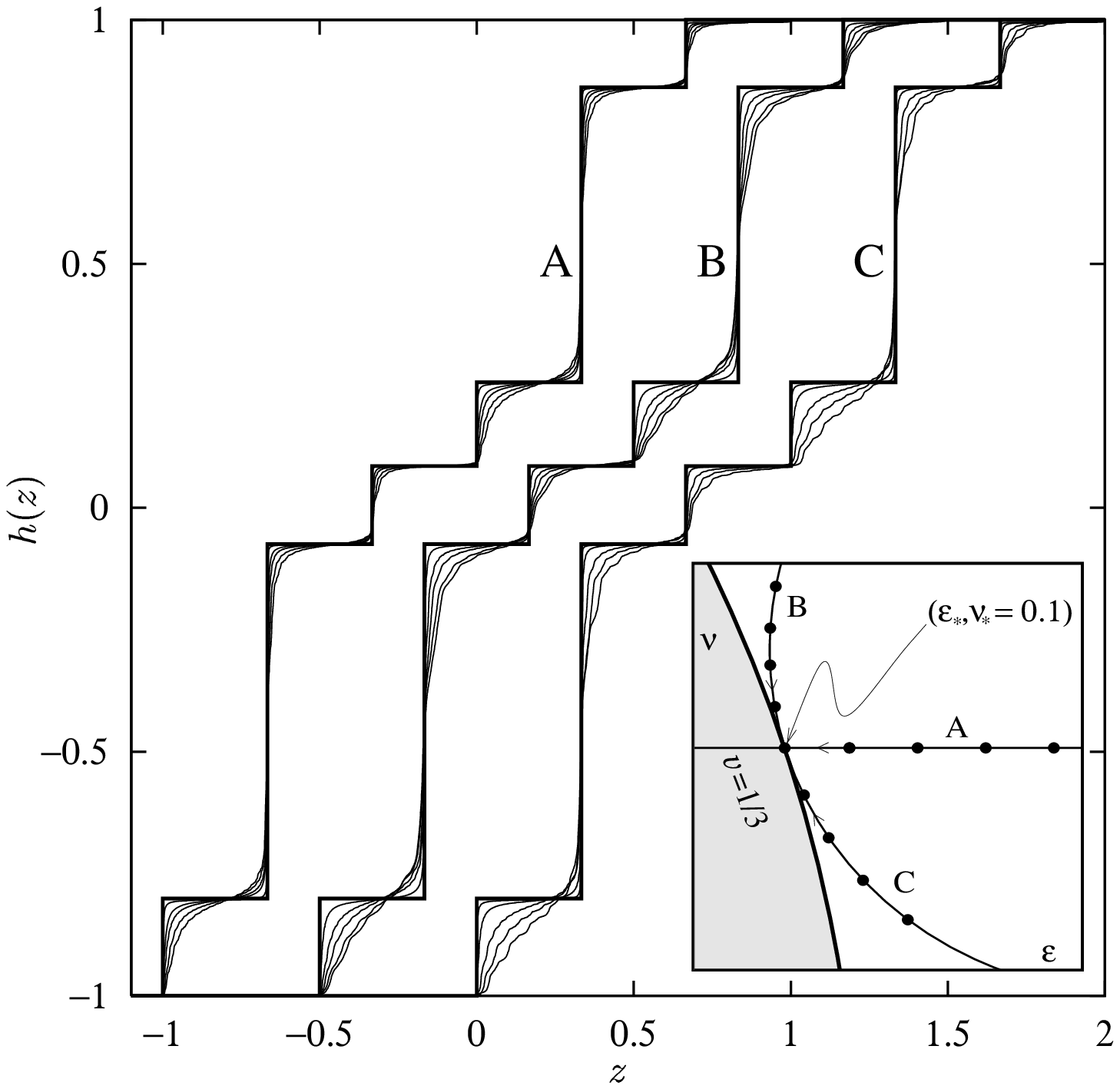}{\FigSize}{\BOUNDARYCAP}

In the next section, we shall explain this phenomenon in 
terms of the dynamics of a one-dimensional map ---the
{\em auxiliary map\/} $\Phi$--- which we now define. 
The idea is to describe the evolution of any site in the 
front by means of a single site, the {\em central site\/} 
$\bar x_t(0)$, defined as the site which is closest to the 
unstable point $x^*$.
The position of the central site moves along the lattice with an
average velocity $v(\e)$, since it follows the centre of the interface.
Following \cite{rcg:modloc,rcg:lowdim}, we define the map $\Phi$\/ as 
\begin{equation} \label{eq:AuxiliaryMap}
\bar x_{t+1}(0)=\Phi(\bar x_t(0)).
\end{equation}

If the velocity is {\em irrational}, the domain of definition of
the map is a set of points dense in an interval (see next section), 
and the possibility exists of extending $\Phi$ continuously to 
the interval.
In Figure \ref{auxmap1G.ps}(a) and (b), we plot the graph of $\Phi$
for a one-way and a diffusive CML, respectively.
The auxiliary map corresponds to the square region depicted
with thick lines,
while the other regions represent delay Poincar\'e maps of 
some neighbouring sites.
Indeed for each neighbour $j$ of the central site, there is
a corresponding auxiliary circle map $\Phi_j$, such that
$\bar x_{t+1}(j)=\Phi_j(\bar x_t(j))$, with $\Phi=\Phi_0$
(see below).

\twoFIG{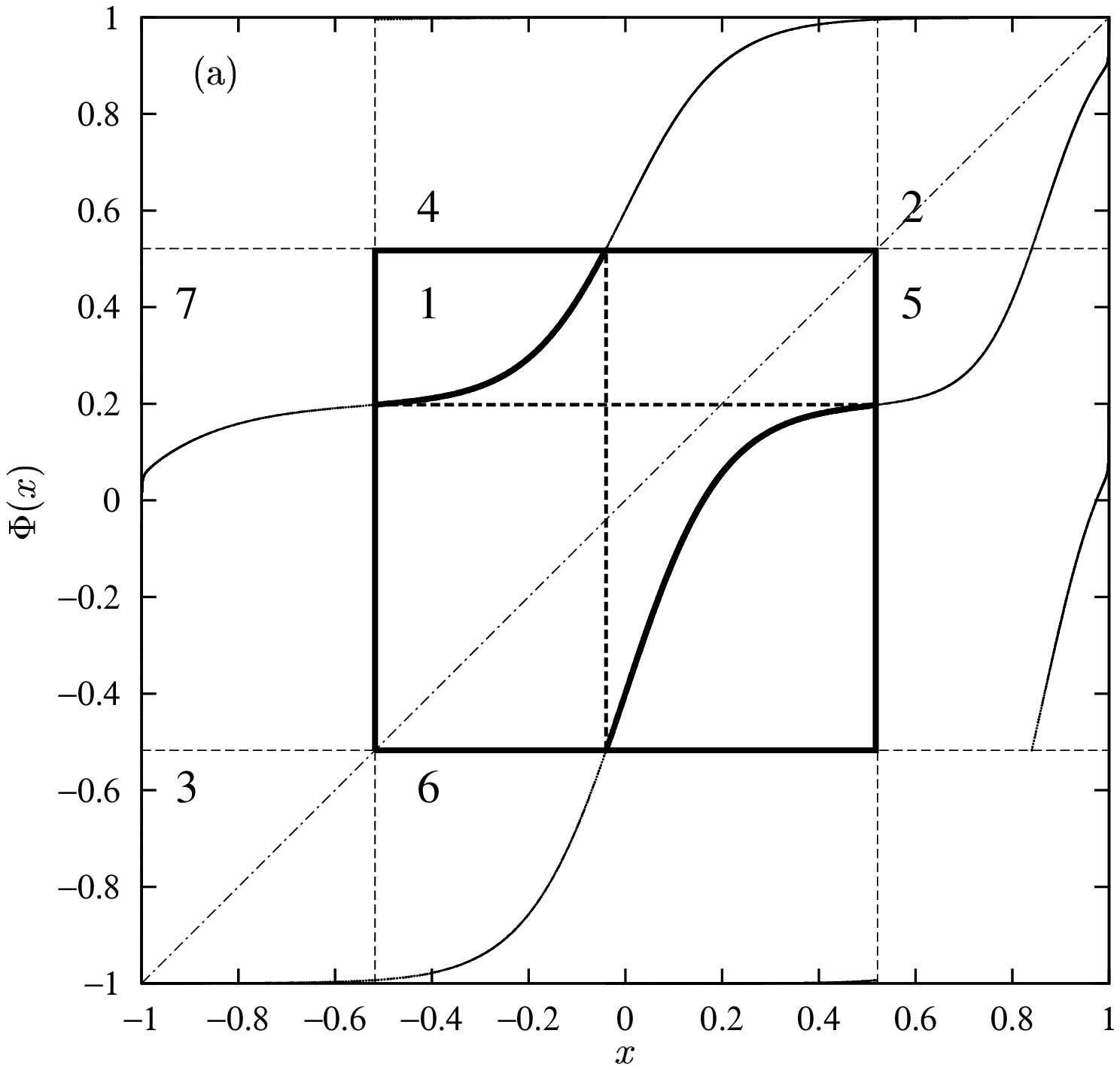}{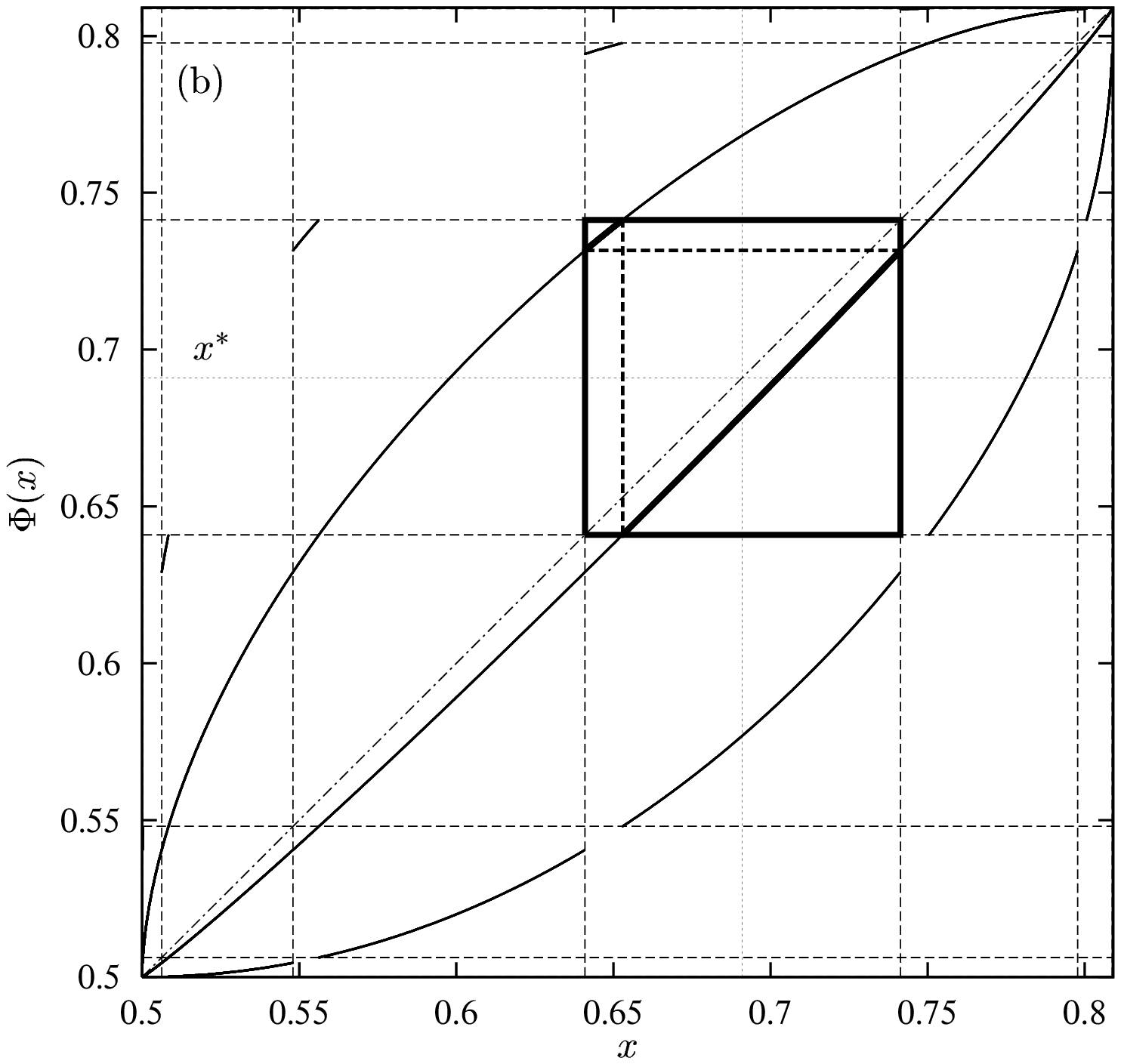}{\FigSize}{\AUXMAPCAP}

If the velocity is {\em rational}, equation (\ref{eq:AuxiliaryMap}) 
defines $\Phi$ only at a finite set of points, and to extend
the domain of definition, one must make use of equation 
(\ref{eq:AuxiliaryMap}) on suitable transients.
We have verified numerically that when a front is perturbed,
the perturbation relaxes quickly onto a one-dimensional 
manifold, along which the original front is approached.
The process of randomly disturbing the front amounts to a random
walk path reconstruction of the one-dimensional manifold.
Such one-dimensional transients were found to be independent 
of the detail of the perturbation, giving a unequivocal 
definition of the auxiliary map also in the rational case.
This is illustrated in Figure \ref{auxmapTongueG.ps}.
Crucially, this construction yields a map that changes 
continuously within the tongue, matching the the behaviour 
at the boundary of the tongue.
Thus we conjecture that the auxiliary map $\Phi$ depends 
continuously on the coupling parameter $\e$.
In the next section we shall explore some consequences
of the continuity.

\oneFIG{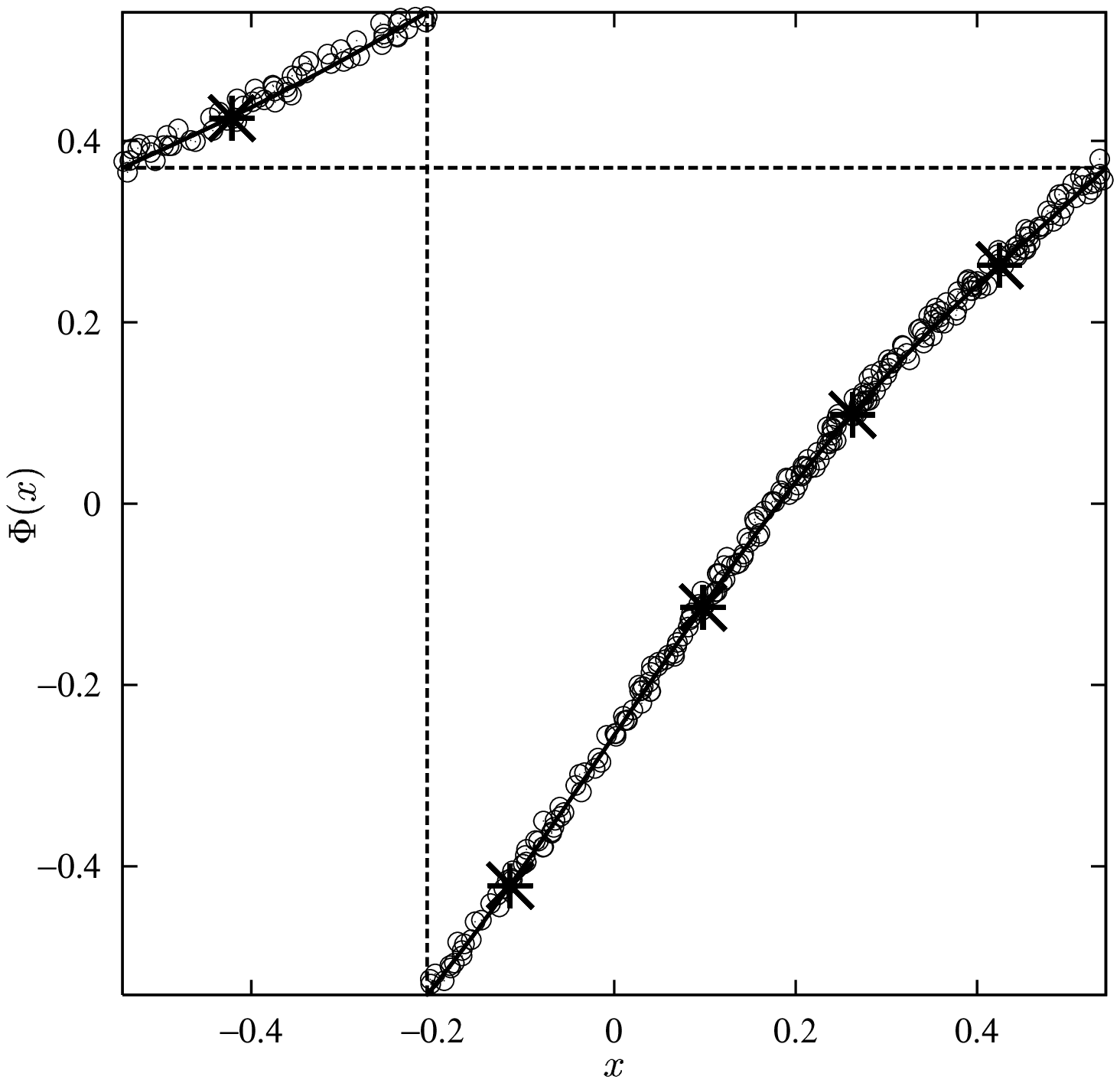}{\FigSize}{\AUXMAPTONGUECAP}

We finally relate the dynamics of the entire front to that 
of the central site, governed by $\Phi(x)$.
Let $\bar x_t(j)$ denote the $j$-th neighbouring site of the
central site $\bar x_t(0)$, where $j$ is positive (negative) 
for the right (left) neighbours.
The dynamics of $\bar x_t(j)$ can be deduced from that of 
$\bar x_t(0)$ and the knowledge of $h$, as follows
\begin{equation}\label{hh-1}
\bar x_t(j)=h\circ \tau_j \circ h^{-1}(\bar x_t(0))
\end{equation}
where $\tau_j$ is the translation by $j$ on $\R$.
Since $\Phi_j(x)$ maps $\bar x_t(j)$ to $\bar x_{t+1}(j)$,
the pair $\left(\bar x_t(j),\,\bar x_{t+1}(j)\right)$ belongs 
to the graph of $\Phi_j$.
By applying the operator $h\circ \tau_j \circ h^{-1}$ to the
function $\Phi(\bar x_t(0))$ we obtain:
$$
\begin{array}{rcl}
h\circ \tau_j \circ h^{-1}\Phi(\bar x_t(0))
&=& h\circ \tau_j \circ h^{-1} (\bar x_{t+1}(0))\\[2.0ex]
&=& \bar x_{t+1}(j) \,=\, \Phi_j\left(\bar x_t(j)\right),
\end{array}
$$
where we used equation (\ref{hh-1}) which relates neighbouring sites.
Thus $h\circ\tau_j \circ h^{-1}$ provides a conjugacy between
$\Phi$ and $\Phi_j$ and enables
us to  reconstruct the whole interfacial
dynamics from the behaviour of the central site.


\section[Mode-locking of the propagation velocity]
{Mode-locking of the propagation velocity{\label{SEC:MODLOC}}}

In this section we show that the auxiliary map $\Phi$ is a 
circle homeomorphism (see Figure \ref{circle.eps}). 
The mode-locking of the front velocity will then follow
from the mode-locking of the rotation number of $\Phi$.
Furthermore, the conjectured continuous dependence of $\Phi$
on $\e$ implies a continuous dependence of the
rotation number on $\e$, and in particular, $\Phi$ takes 
all rotation numbers between any two realised values. 
For instance, the front velocity in a one-way CML takes the values
$0$ and $1$ for $\e=0$ and $1$, 
respectively, and thus as the coupling parameter varies,
all velocities $v\in [0,1]$ are realised.
For a diffusive CML only an interval $[0,v_{max}]$ is
attained since the maximum velocity $v_{max}=v(\e=1)$ 
does not reach 1 because of the competition between the attractors.

\oneFIG{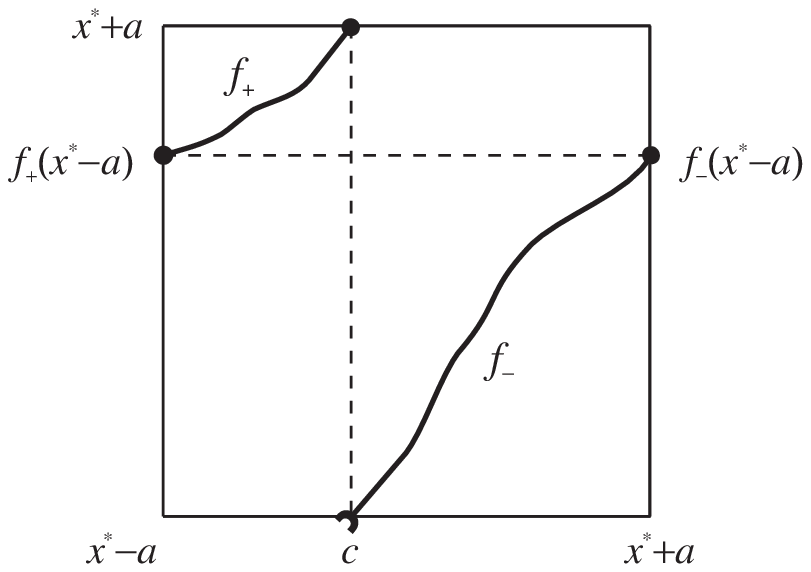}{\FigSize}{\CIRCLECAP}

Let us consider a continuous and increasing travelling front 
$h(i-vt+i_0)$ with positive irrational velocity $0<v<1$.
The largest possible separation between $\bar x_t(0)$ and $x^*$
corresponds to the position of $h$ for which two consecutive 
points on the lattice are equally spaced from the unstable point $x^*$.
Suppose that the front shape $h$ is positioned such that for
site $i$, we have $h(i)=x^*$. 
We choose $\alpha$ such that
\begin{equation}\label{h_alpha}
h(i-\alpha) = x^* - a \quad {\rm and} \quad
h(i+1-\alpha) = x^* + a
\end{equation}
where $0\leq a \leq\min(|x_+^*-x^*|,|x_-^*-x^*|)$.
By adding the two equations in (\ref{h_alpha}) one obtains
an equation for $\alpha$, and $a$ can then be evaluated.
If the front is at a position where it satisfies the equations 
(\ref{h_alpha}) for some $i$, then the $i$-th and $(i+1)$-th 
sites are equally spaced from $x^*$, and the dynamics of the site 
closest to $x^*$ is contained in the interval $[x^*-a,x^*+a]$.
Any shift of the front will cause either one of the two 
sites to be closer to $x^*$ than originally. 

We now follow the dynamics of $\bar x_t(0)$ in $[x^*-a,x^*+a]$.
Suppose that at time $\tau$ the $i$-th site is the closest to $x^*$ so
$\bar x_\tau(0) = x_\tau(i)$. We want to know which site will be 
closest to $x^*$ at time $\tau+1$. Since we are considering
the case $v>0$ there are two possibilities:
a) the $i$-th site again ($\bar x_{\tau+1}(0)=x_{\tau+1}(i)$)
or b) the $(i+1)$-th site ($\bar x_{\tau+1}(0)=x_{\tau+1}(i+1)$).
Redefining $h_t(i)=h(i-vt+i_0)$, we find two cases
\begin{equation}\label{h1}
h_{\tau+1}^{-1}(\bar x_{\tau+1}(0))=\left\{
\begin{array}{ll}
h_\tau^{-1}(\bar x_\tau(0)) &\quad\mbox{(a)}\\[2.0ex]
h_\tau^{-1}(\bar x_\tau(0))+1 &\quad\mbox{(b)}
\end{array}
\right. .\end{equation}
But, by definition, $h_{\tau+1}(x)=h_{\tau}(x-v)$, so from equation
(\ref{h1}) one obtains
\begin{equation}\label{h2}
\bar x_{\tau+1}(0)=\left\{
\begin{array}{ll}
f_-(\bar x_\tau(0)) &\quad\mbox{(a)}\\[2.0ex]
f_+(\bar x_\tau(0)) &\quad\mbox{(b)}
\end{array}
\right. \end{equation}
where
\begin{equation}\label{h3}
\begin{array}{rcl}
f_-(x) & = & h_\tau\left(h_\tau^{-1}(x)-v\right)\\[2.0ex]
f_+(x) & = & h_\tau\left(h_\tau^{-1}(x)-v+1\right).
\end{array}\end{equation}
The functions $f_-$ and $f_+$ inherit some of the properties of $h$. 
In particular, $f_-$ and $f_+$ are continuous and increasing.
In the the interval $[x^*-a,x^*+a]$ we have that $f_-(x)<f_+(x)$, 
because $h$ is increasing, so we just evaluate at the following points:
$$
\begin{array}{rcl}
f_-(x^*+a)&=&h_\tau(h_\tau^{-1}(x^*+a)-v)\\[1.0ex]
          &=&h_\tau(i+1-\alpha-v)\\[3.0ex]
f_+(x^*-a)&=&h_\tau(h_\tau^{-1}(x^*-a)-v+1)\\[1.0ex]
          &=&h_\tau(i-\alpha-v+1)
\end{array}
$$
where we have made use of equations (\ref{h_alpha}).
Thus we have the periodicity condition
\begin{equation}\label{h5}
f_-(x^*+a)=f_+(x^*-a).
\end{equation}
Next we find when $f_-$ and $f_+$ reach the extrema of the interval 
$[x^*-a,x^*+a]$.
To this end we determine $c_\pm$ such that $f_\pm(c_\pm)=x^*\pm a$.
So we solve
$$
\begin{array}{rl}
 &
\left\{ \begin{array}{rcccl}
f_-(c_-)&=&h_\tau(h_\tau^{-1}(c_-)-v)&=&x^*-a\\[2.0ex]
f_+(c_+)&=&h_\tau(h_\tau^{-1}(c_+)-v+1)&=&x^*+a
\end{array}\right. \\[5.0ex]
\Rightarrow &
\left\{ \begin{array}{rcccl}
h_\tau^{-1}(c_-)-v&=&h_\tau^{-1}(x^*-a)&=&i-\alpha\\[2.0ex]
h_\tau^{-1}(c_+)-v+1&=&h_\tau^{-1}(x^*+a)&=&i+1-\alpha
\end{array}\right. \\[6.0ex]
\end{array}
$$
whence $h_\tau^{-1}(c_-)= h_\tau^{-1}(c_+)$, and since $h$ 
is monotonic, we have that $c_-= c_+=c.$

Therefore, the map $\Phi$ giving the dynamics of the central
site (\ref{eq:AuxiliaryMap}) is given by
\begin{equation}\label{Phi}
\Phi(x)=\left\{
\begin{array}{lrcccl}
f_+(x) & \mbox{~~if~~} x^*-a &\leq &x&\leq& c \\[2.0ex]
f_-(x) & \mbox{~~if~~} x^*+a &\geq &x& >  & c.
\end{array}
\right.
\end{equation}
From the above properties of $f_-$ and $f_+$, it follows that 
the auxiliary map $\Phi$ is a homeomorphism of the circle 
(see Figure \ref{circle.eps}).

A natural binary symbolic dynamics for $\Phi$ is introduced
by assigning the symbols `0' and `1' whenever the branch 
$f_-$ or $f_+$, respectively, is used in (\ref{h1}).
These symbols corresponds to the central site $x(i)$ remaining 
unchanged, or being replaced by the new site $x(i+1)$,
respectively.

\oneFIG{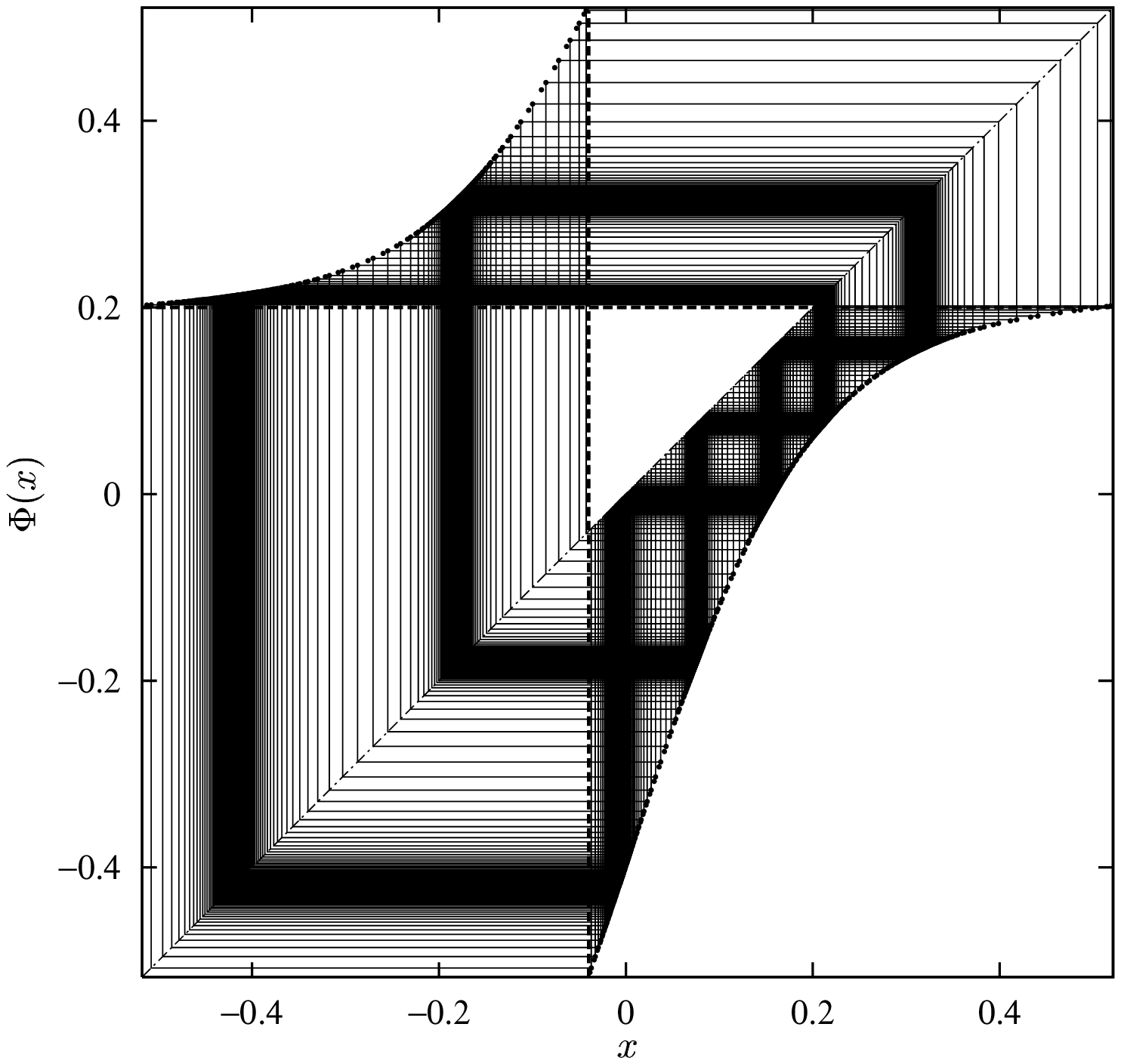}{\FigSize}{\INTERMITTENCYCAP}
 
Every time a `1' is encountered, the front advances by roughly 
one site. So the density of `1's in the sequence gives an approximation
to the velocity, which becomes exact in the limit $t\rightarrow\infty$.
In terms of the circle map, the proportion of `1's in the sequence
corresponds to its rotation number $\rho$:
\begin{equation}
\rho(\e)=v(\e)
=\lim_{t\rightarrow\infty}{{\ds 1\over \ds t}\sum_{i=1}^{t}{s_i}},
\end{equation}
where $s_i$ is the $i$-th term in the symbolic sequence.
We have stressed the $\e$-dependence of $\rho$, since for a fixed
local map, $\Phi$ depends on $\e$, and so does its rotation number.
Because all sites $\bar x(j)$ belong to the same front, the
site interchanges all occur at the same time, and therefore the
rotation number of any $\Phi_i$ is the same as the one for $\Phi$.

The representation of the motion of a front as a circle map
implies the likelihood of mode-locking for rational velocities,
corresponding to Arnold tongues in parameter space, and it
affords a simple explanation of the various dynamical 
phenomena described in the previous sections.

The appearance of a $q$-period tongue as $\e$ is varied
thorough some critical value $\e_*$, corresponds to a 
fold bifurcation of $\Phi^q$.
Generically, a pair of period-$q$ orbits is created at $\e=\e_*$. 
Thus the orbits of $\Phi$ will undergo intermittency in the region 
of the period-$q$ orbit for $\e_n$ close to $\e_*$.
The intermittency will manifest itself in the graph of $\Phi$ as 
shown by the darkly shaded areas of the orbit web in Figure 
\ref{intermittencyG.ps}. 

Moreover, the periodic orbit will form towards the centre of the dark 
bands and the corresponding front shape will ``flatten'' at the heights 
taken by the periodic points because of the time spent in their 
neighbourhood by the orbits of $\Phi$ for $\e_n\approx \e_*$.  
It then follows that the approximating fronts will form steps for the 
periodic front with the periodic points close to their centre points,
and independently from the parametric path chosen to approach the
boundary point (see Figure \ref{boundary.ps}).

\oneFIG{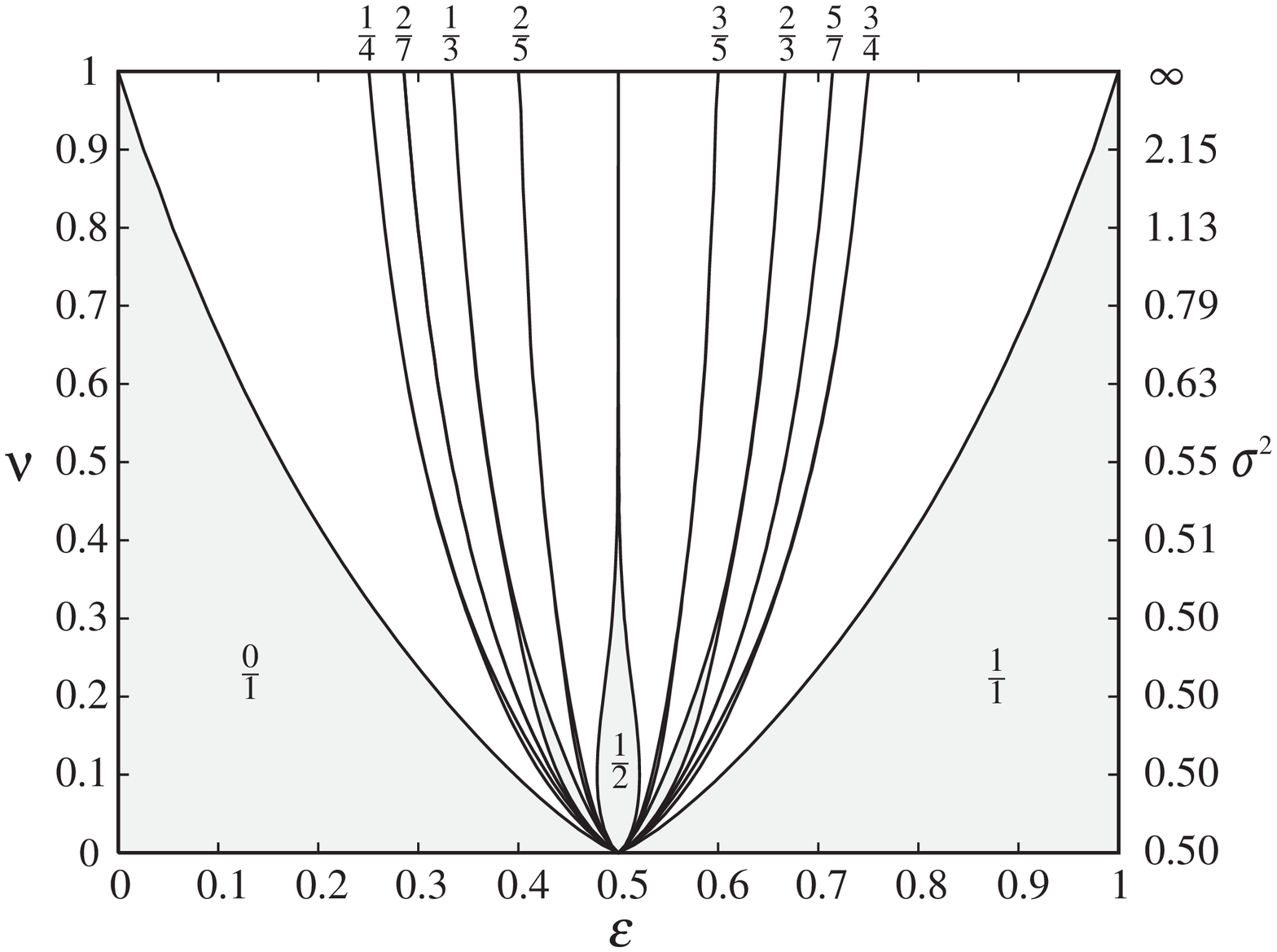}{\FigSize}{\TONGUESCAP}

In Figure \ref{tongues.eps} we plot
the main mode-locking regions in parameter space
(Arnold tongues), corresponding to $v=p/q$ with small $q$.
Here the local map is given by $f(x)=\tanh(x/\nu)$, while
the parameters vary within the unit square:  $(\e,\nu)\in[0,1]^2$.
We believe that mode-locking is a common phenomenon in front
propagation in CMLs, because the nonlinearity of the local map
induces nonlinearity in the auxiliary map \cite{rcg:modloc,rcg:lowdim},
and mode-locking is generic for such maps.
However, this phenomenon often takes place on very small parametric 
scales, since the width of the tongues decreases sharply
with increasing $\nu$ (Figure \ref{tongues.eps}).
This explains why this phenomenon has not been widely reported
(with the notable exception of the large $v=0$ region, corresponding
to the well-known propagation failure in the anti-continuum limit
\cite{Aubry:90-MacKay:95}).

\oneFIG{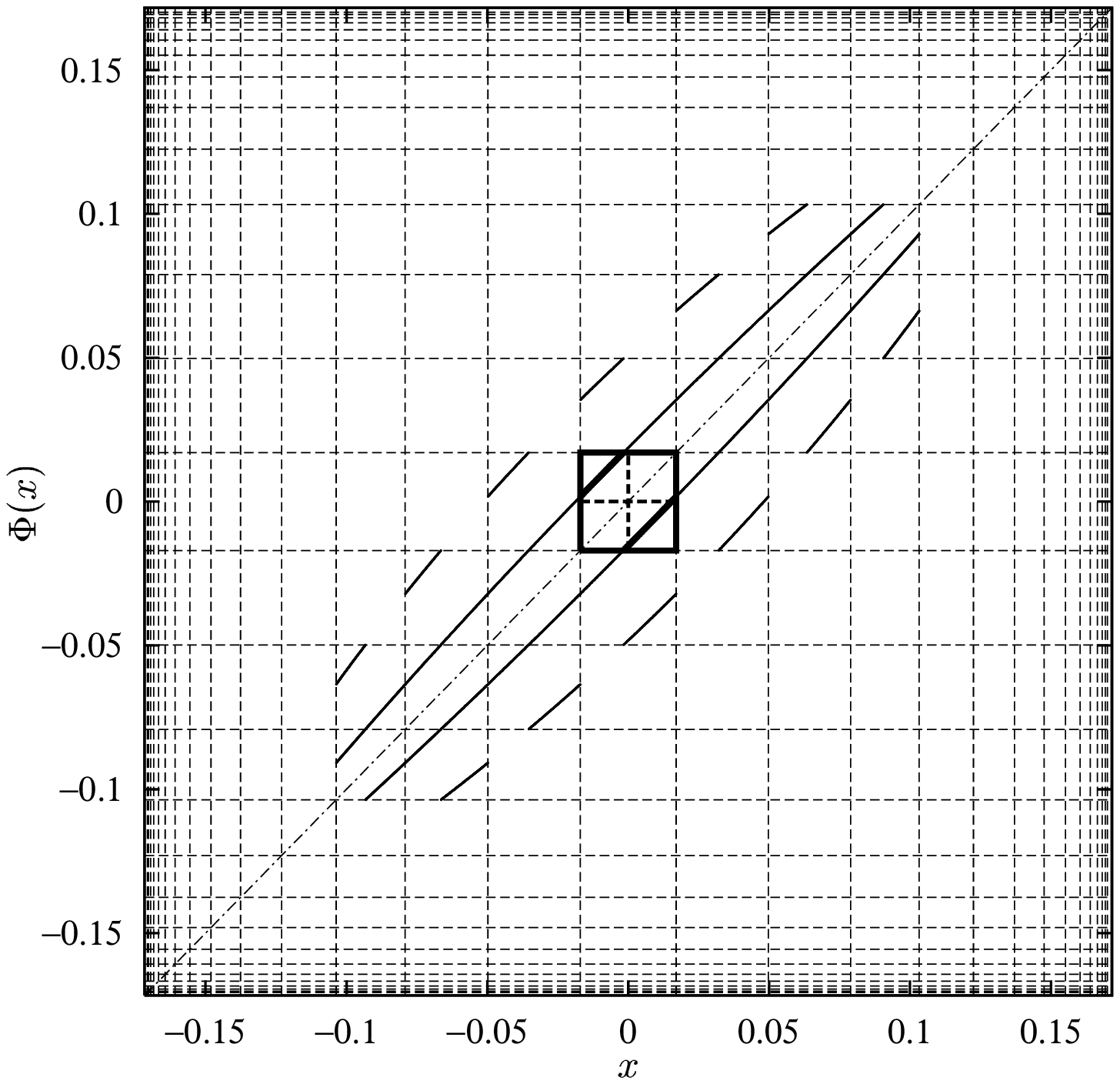}{\FigSize}{\AUXMAPCONTINUUMCAP}

In the continuum limit (see Figure \ref{tongues.eps}), the stability
of the attractors $x_\pm^*$ becomes weaker, causing the front to broaden.
In Figure \ref{auxmapContinuumG.ps} we plotted the auxiliary maps $\Phi_i$
corresponding to $\nu=100/101 \simeq 1$ for the one-way CML with
local map $f(x)=\tanh(x/\nu)$. This figure should be compared with
Figure \ref{auxmap1G.ps}, corresponding to a narrower front.
The domain of each $\Phi_i$ is now smaller, since the interval
$I=[x_-^*,x_+^*]$ has to be shared between a larger number of sites.
As a consequence, the nonlinearity of each $\Phi$ is reduced
(note that the auxiliary maps in Figure \ref{auxmapContinuumG.ps} 
are almost linear) and with it the size of the tongues.
Thus, the larger the width $\sigma^2$ of the travelling front, the 
thinner the mode-locking tongue (see right hand side scale in
Figure \ref{tongues.eps}).

\section*{Acknowledgments}
RCG would like to acknowledge DGAPA-UNAM (M\'exico) for the financial
support during the preparation of this paper. This work was partially
supported by EPSRC GR/K17026.


\end{document}